\documentclass{iau} 
\usepackage{graphicx}
\usepackage{hyperref}

\title[Sunspot group tilt angles] 
{Sunspot group tilt angles from drawings for cycles 19-24 }

\author[E. I\c{s}\i k, S. I\c{s}\i k \& B. B. Kabasakal]   
{Emre I\c{s}\i k$^{1,2},$
Seda I\c{s}\i k$^3$ \and Bahar B. Kabasakal$^2$}

\affiliation{$^1$Max-Planck-Institut f\"ur Sonnensystemforschung, Justus-von-Liebig-Weg 
3, 37077, G\"ottingen, Germany \\ email: {\tt isik@mps.mpg.de} \\[\affilskip]
$^2$Feza G\"ursey Center for Physics and Mathematics, Bo\u{g}azi\c{c}i University, 
Kuleli 34684 Istanbul, Turkey \\[\affilskip]
$^3$Kandilli Observatory and Earthquake Research Institute, Bo\u{g}azi\c{c}i University,
Kuleli 34684 Istanbul, Turkey} 
\pubyear{2018}
\volume{340}  
\jname{Long-Term Datasets for the Understanding of Solar and Stellar Magnetic Cycles}
\editors{D. Banerjee, J. Jiang, K. Kusano \& S.K. Solanki, eds.}
\begin{document}

\maketitle

\begin{abstract}
The tilt angle of a sunspot group is a critical quantity in the surface transport of 
magnetic flux and the solar dynamo. To contribute long-term databases 
of the tilt angle, we developed an IDL routine, which allows the user to 
interactively select and measure sunspot positions and areas on the solar disc. 
We measured the tilt angles of sunspot groups for solar cycles 19-24 (1954.6-2017.8), 
using the sunspot drawing database of Kandilli Observatory. 
The method is similar to that used in the discontinued Mt. Wilson and Kodaikanal databases, 
with the exception that sunspot groups were identified manually, 
which has improved the accuracy of the resulting tilt angles.  
We obtained cycle averages of the tilt angle and compared them with the values from 
other datasets, keeping the same group selection criteria. We conclude that 
the previously 
reported anti-correlation with the cycle strength needs further investigation. 
\end{abstract}

\firstsection 
\section{Introduction}
The angle between the line joining opposite 
polarities and the local latitude is called the tilt angle of a sunspot group. 
It is defined as positive when the preceding polarity is 
closer to the equator. This is an important quantity because, through latitudinal separation 
between opposite polarities of bipolar regions, it controls the net magnetic flux which 
contributes to the global 
axial dipole moment of the Sun, which is well-correlated with the strength of the subsequent 
activity cycle. Systematic changes in the tilt 
angle by inflows at the surface \cite{jiang10} or by stabilisation of deep-seated flux tubes 
\cite{isik15} can play important roles in the nonlinear saturation of the solar dynamo. 

It is evident from the literature that long-term tilt angle databases should be enriched, to 
evaluate the relationship of its cycle averages with amplitude fluctuations of solar cycles. 
Our main motivation to construct an additional dataset is to make an independent test of the 
reported anti-correlation between cycle-averaged tilt angle and the cycle strength \cite{dasi10}. 

\begin{figure}
\hskip2cm{(a)}
\vskip-7mm
\centerline{\includegraphics[width=.6\columnwidth]{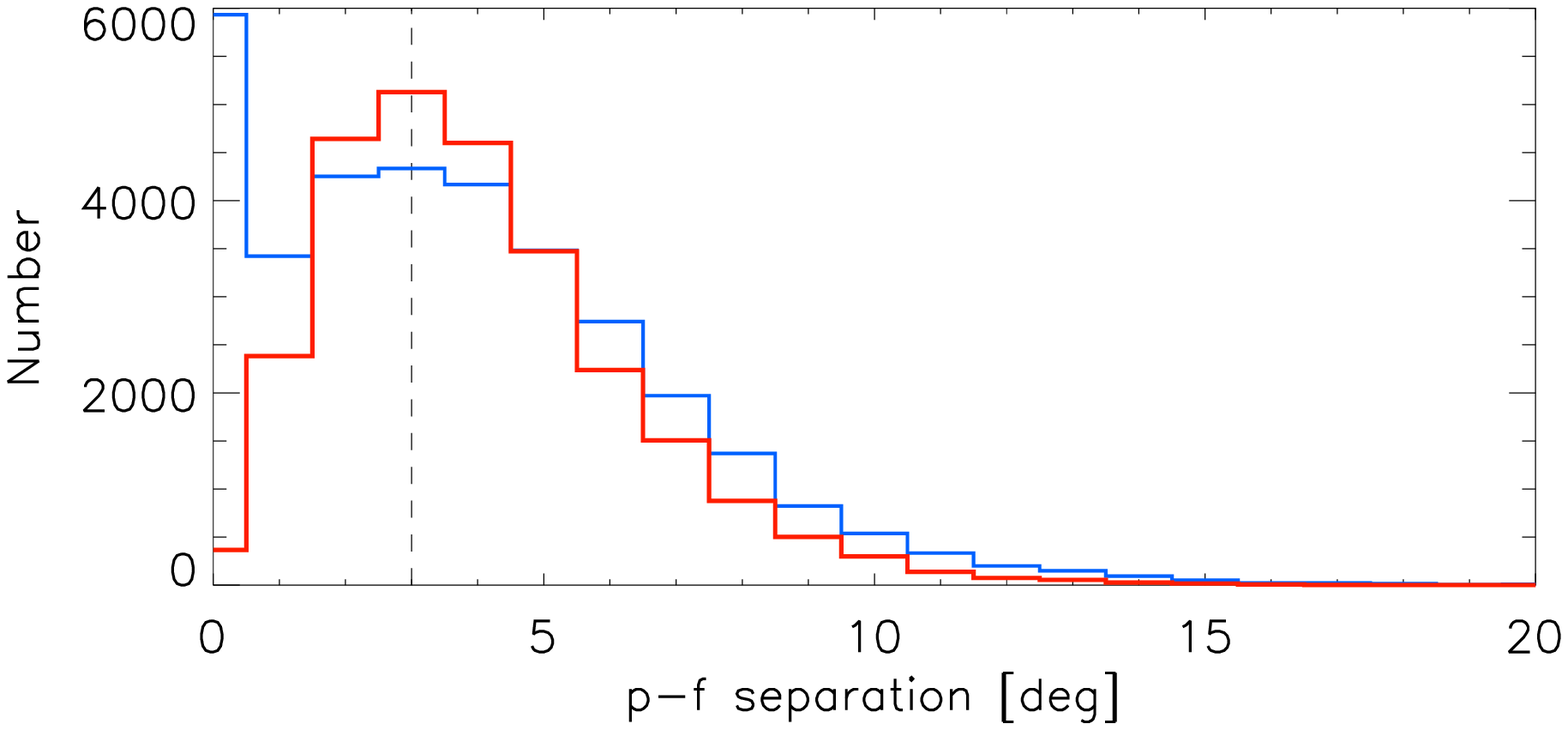}}
\hskip2cm{(b)}
\vskip-7mm
\centerline{\includegraphics[width=.6\columnwidth]{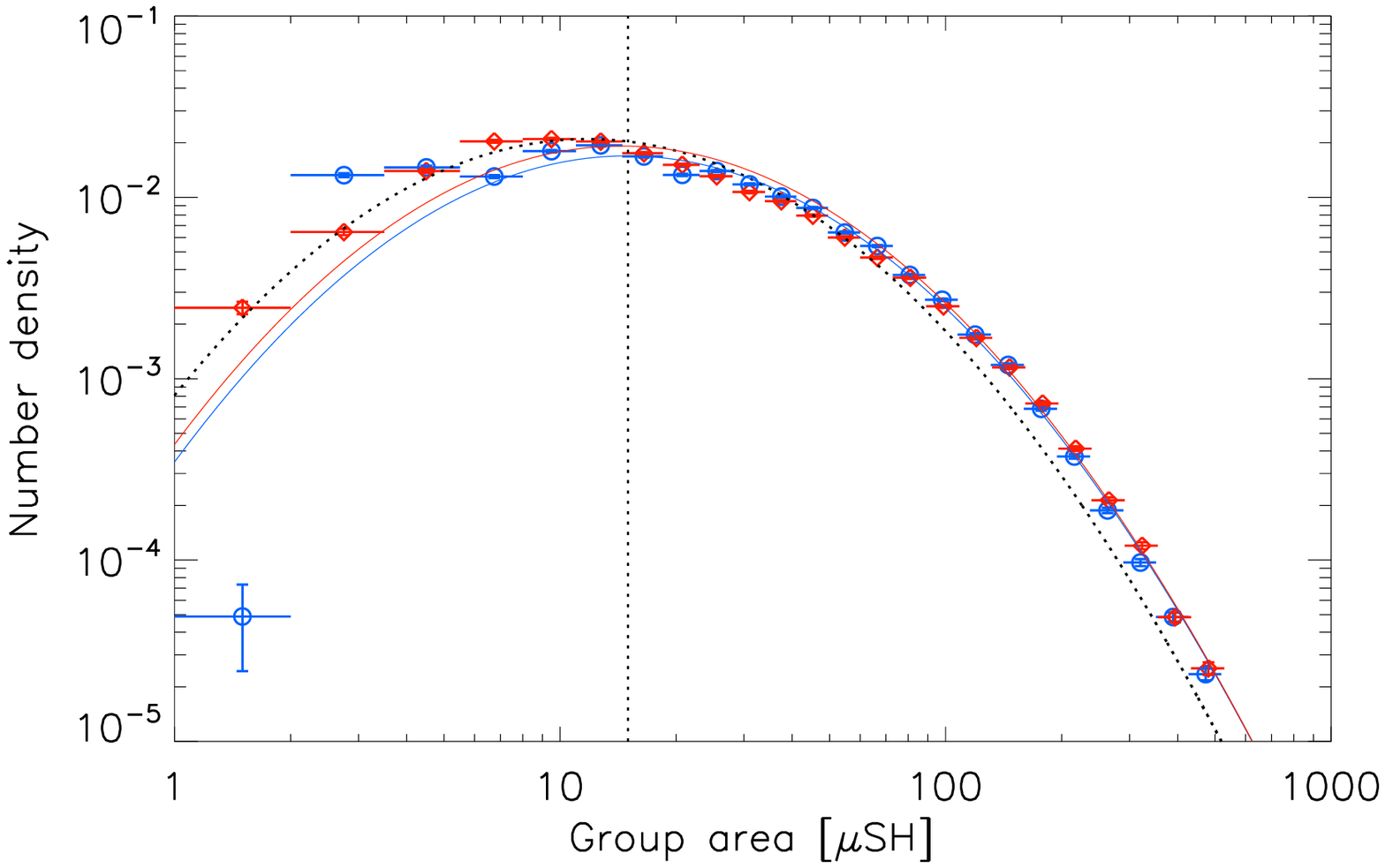}}
 \caption{Distributions for (a) the angular separation between p- and f- parts, and 
 (b) the group area for $S>3^\circ$, the vertical line in panel (a), 
   normalised to the number of groups with $A>1$~MSH (Kandilli in red, DPD in blue). 
 The vertical error bars show the standard deviation of the mean, and the horizontal bars show the area bins. 
 Lognormal fits were made for $A>15~\mu$SH, and a model function from 
 \cite{bs05} is also shown (dotted curve).}
   \label{fig:separea}
\end{figure}

\section{Measuring sunspot groups}
We used the digitised web archive of Kandilli Observatory sunspot drawings, which 
have been systematically produced since 1944, using a 20-cm equatorial-mount refractor 
with 307-cm focal length. We developed and used an interactive IDL routine, to record 
the positions and areas of umbral features, including pores, so far as they were 
detected and drawn by the observers. The selection of pixels is based on 
a region-of-interest algorithm. We also measured the inclination of the solar axis relative 
to the local meridian, at the time of each observation. This is used not only to 
determine the position angle of each spot, but also to correct for the small tilt of the 
scanned drawing, by referring to the exact value of the inclination. 
To measure the tilt angle, we use the method of \cite{howard91}, with the exception 
that we adopt group identifications of the observers, rather than using an automated algorithm. There are 1-2 year data gaps in parts of cycle 19 (1955-57 and 1962-63) 
and in cycle 22 (1993). 

Figure~\ref{fig:separea}a shows the histogram of the angular separation, $S$, 
between the preceding 
(p-) and follower (f-) parts of our groups with area $A>1~\mu$SH (micro solar hemisphere) 
and distance to disc centre, $\rho<60^\circ$.
For comparison, we plotted the Debrecen Photoheliographic Record 
(DPD\footnote{\url{http://fenyi.solarobs.csfk.mta.hu/}}) results \cite{dpd16} keeping the 
same criteria. Since DPD provides the
angular distance to the central meridian (CMD), we took CMD$<60^\circ$. 
For DPD, we chose only those group measurements which have a measured tilt angle. 
A local peak around $3^\circ$ is present in both datasets. In our data,  
the distribution drops rather sharply towards more compact groups, owing to the tendency of 
observers not to classify them as bipolar (we measured only `bipolar' regions, excluding 
McIntosh types A and H). 
In DPD data, however, there is an additional peak at the smallest 
separations (see also \cite[Baranyi 2015]{baranyi15}, Fig.~3). To have a reliable comparison 
of tilt angles in these datasets, we considered only groups with $S>3^\circ$, as also 
suggested by \cite[Baranyi (2015)]{baranyi15} and \cite{pavai15}, in addition to the criteria 
described for Fig.~\ref{fig:separea}a. 
The discrete area distributions for both datasets are shown in Fig.~\ref{fig:separea}b, 
along with lognormal function fits, which are applied to data above a group area of $15$~MSH. We 
adopted this threshold from \cite{bs05}, whose model distribution including lognormal
spot decay rates is also shown for comparison. 
Our data fits well with the DPD in terms of the group area distribution, except for very 
small areas of a few $\mu$SH, for which intrinsic uncertainties become comparable with the group area in any case. 

\section{Joy's law and cycle-averaged tilt angles}
\begin{figure}
\centerline{\includegraphics[width=.8\columnwidth]{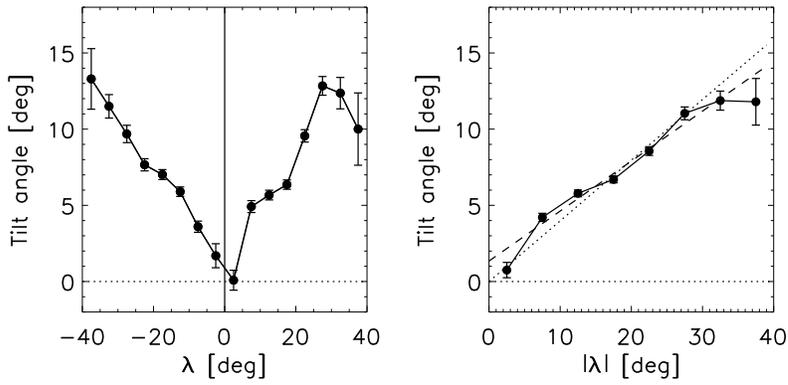}}
 \caption{Joy's law for signed (left panel) and unsigned (right panel) latitude. The 
 error bars show standard deviation of the means over $5^\circ$ bins.}
   \label{fig:joy}
\end{figure}

Figure~\ref{fig:joy} shows the latitude dependence of the tilt angle of Kandilli data, 
for different hemispheres and for unsigned latitude. We applied here the same group 
selection criteria as in Fig.~\ref{fig:separea}b. 
We confirm \cite[Baranyi's (2015)]{baranyi15} result that there is a persistent pattern in 
the latitudinal dependence of the tilt angle. In particular, there is a significant 
north-south asymmetry in Joy's law for our six-cycle averages, and a double-hump pattern 
in the global averages of $\alpha(|\lambda|)$, all very similar to those found for three cycles 
by \cite[Baranyi (2015, Figs.~8-9)]{baranyi15}. A linear least-squares fit gives a Joy's law 
slope of $0.398\pm 0.006$, where the y-intercept is forced to zero. 
When the fit is confined to $|\lambda|<30^\circ$ we find $0.405\pm 0.007$, 
which is closer to that found by \cite[Baranyi (2015)]{baranyi15}. 
When the intercept is not forced, 
we find $\alpha=(0.328\pm 0.015)\lambda+(1.34\pm 0.27)$. 

\begin{figure}
\centerline{\includegraphics[width=.8\columnwidth]{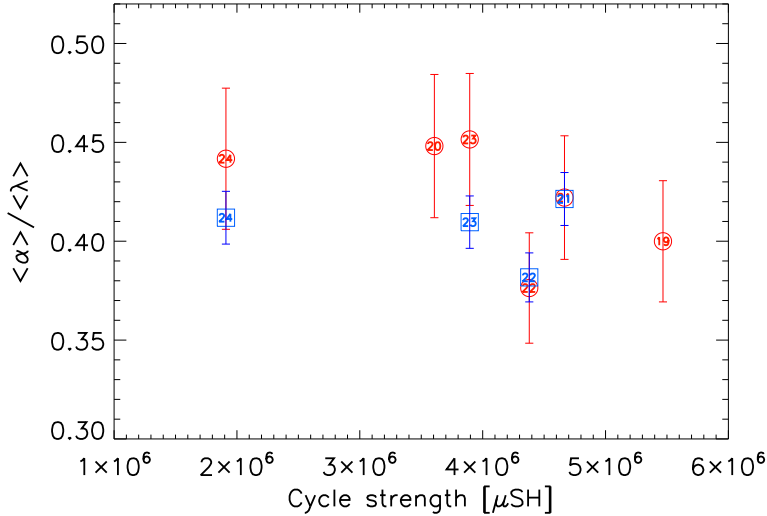}}
 \caption{Cycle-averaged tilt angle normalised to cycle-averaged latitude, as a function of 
   cycle strengths, for Kandilli (red circles) and DPD (blue squares) datasets. 
   Note that cycle 24 is covered until 2017.8 in Kandilli, and 2016.0 in DPD. 
 }
   \label{fig:anticorr}
\end{figure}

We calculated cycle-averages of the tilt angles $\langle\alpha\rangle$ and 
latitudes $\langle\lambda\rangle$, of all groups with the 
same criteria as in Fig.~\ref{fig:separea}b to check for a systematic dependence on 
cycle strength. 
Figure~\ref{fig:anticorr} shows the ratio $\langle\alpha\rangle/\langle\lambda\rangle$ 
for Kandilli and DPD. 
It is known that unipolar and compact groups have an almost zero mean tilt angle, whereas ephemeral regions have negative mean tilt angles in magnetographic results 
\cite{tlatov13}. Owing to the lowest-separation constraint, such groups are now eliminated. 
This is the reason for the systematic differences between $\langle\alpha\rangle$ (thus $\langle\alpha\rangle/\langle\lambda\rangle$) of our data 
and those in \cite[Dasi-Espuig \etal\ (2010)]{dasi10}. The correlation 
coefficient we find for Kandilli data with 6 cycles is $-0.57$ with a $p$-value of $0.23$ 
for $S>3^\circ$ (Fig.~\ref{fig:anticorr})
and $-0.75$ with a $p$-value of $0.08$ when $S>2.5^\circ$. In the latter case, the 
discrepancy in Cycle 23 between the two datasets also decreases. 

\section{Conclusion}
We have found a remarkable similarity between our data covering cycles 1954-2017 and 
DPD, which provides sunspot group tilt angles for 1973-2016. 
This demonstrates the great potential of sunspot drawing archives in deriving important 
spatio-temporal information on sunspot group properties.  
Despite differences in the tilt angle determination, 
both datasets compared here have the common feature of careful group identification and the elimination 
of unipolar and/or very compact, complex groups. Our results do not indicate a 
clear anti-correlation between $\langle\alpha\rangle/\langle\lambda\rangle$ and the cycle 
strength, contrary to \cite[Dasi-Espuig \etal\ (2010, 2013)]{dasi10,dasi13}. In a subsequent 
paper, we will present our work in more detail, also by revisiting Mt. Wilson and Kodaikanal datasets. 

\section*{Acknowledgements}
This work was supported by the Turkish Scientific and Technological Research Council, 
under grant 114F372. EI acknowledges the Young Scientist 
Award Programme BAGEP 2016 granted by the Society of the Science Academy, Turkey.


\begin{thebibliography}{}

\bibitem[(Baranyi 2015)]{baranyi15} Baranyi, T.\ 2015, \textit{MNRAS}, 447, 1857 
\bibitem[(Baranyi et al. 2016)]{dpd16} Baranyi, T., Gy{\H o}ri, L., \& Ludm{\'a}ny, A.\ 2016, \textit{Solar Phys.}, 291, 3081 
\bibitem[Baumann \& Solanki (2005)]{bs05} Baumann, I., \& Solanki, S.~K.\ 2005, \textit{A\&A}, 443, 1061 
\bibitem[(Dasi-Espuig \etal\ 2010)]{dasi10} Dasi-Espuig, M., Solanki, S.~K., Krivova, N.~A., Cameron, R., \& Pe{\~n}uela, T.\ 2010, \textit{A\&A}, 518, A7 
\bibitem[(Dasi-Espuig \etal\ 2013)]{dasi13} Dasi-Espuig, M., Solanki, S.~K., Krivova, N.~A., Cameron, R., \& Pe{\~n}uela, T.\ 2013, \textit{A\&A}, 556, C3 
\bibitem[Howard (1991)]{howard91} Howard, R.~F.\ 1991, \textit{Solar Phys.}, 136, 251 
\bibitem[(I{\c s}{\i}k 2015)]{isik15} I{\c s}{\i}k, E.\ 2015, \textit{ApJ} (Letters), 813, L13 
\bibitem[(Jiang \etal\ 2010)]{jiang10} Jiang, J., I{\c s}{\i}k, E., Cameron, R.~H., Schmitt, D., \& Sch{\"u}ssler, M.\ 2010, \textit{ApJ}, 717, 597 
\bibitem[Senthamizh Pavai \etal\ (2015)]{pavai15} Senthamizh Pavai, V., Arlt, R., Dasi-Espuig, M., Krivova, N.~A., \& Solanki, S.K. 2015, \textit{A\&A}, 584, A73 
\bibitem[(Tlatov \etal\ 2013)]{tlatov13} Tlatov, A., Illarionov, E., Sokoloff, D., \& Pipin, V.\ 2013, \textit{MNRAS}, 432, 2975 

\end{thebibliography}
\end{document}